\documentclass[conference]{IEEEtran}
\IEEEoverridecommandlockouts
\usepackage{cite}
\usepackage{amsmath,amssymb,amsfonts}
\usepackage{algorithmic}
\usepackage{graphicx}
\graphicspath{{images/}}
\usepackage{textcomp}
\usepackage{xcolor}
\usepackage{balance}
\usepackage{url}
\def\BibTeX{{\rm B\kern-.05em{\sc i\kern-.025em b}\kern-.08em
    T\kern-.1667em\lower.7ex\hbox{E}\kern-.125emX}}

\newcommand{\loss}{\mathcal{L}}

\begin{document}

\title{%ARTInp:Inpainted Images for CBCT-to-CT Image Translation in Radiotherapy\\
ARTInp: CBCT-to-CT Image Inpainting and Image Translation in Radiotherapy

% Taking the grant off for anonymization
% \thanks{This work is supported by Italian Ministry of Health (grant number: GR-2019-12370739, project: AuToMI).}
}

%\author{\IEEEauthorblockN{Anonymous Authors}}

\author{
\IEEEauthorblockN{Ricardo Coimbra Brioso\IEEEauthorrefmark{2}, Leonardo Crespi\IEEEauthorrefmark{2},Andrea Seghetto\IEEEauthorrefmark{2}, Damiano Dei\IEEEauthorrefmark{3}\IEEEauthorrefmark{4}, Nicola Lambri\IEEEauthorrefmark{3}\IEEEauthorrefmark{4},\\
Pietro Mancosu\IEEEauthorrefmark{4}, and Marta Scorsetti\IEEEauthorrefmark{3}\IEEEauthorrefmark{4}, Daniele Loiacono\IEEEauthorrefmark{2}}
\smallskip
\IEEEauthorblockA{\IEEEauthorrefmark{1}Dipartimento di Elettronica, Informazione e Bioingegneria, Politecnico di Milano, Milan, Italy}
%\{leonardo.crespi,daniele.loiacono\}@polimi.it, mattia.portanti@mail.polimi.it\\
\IEEEauthorblockA{\IEEEauthorrefmark{3}Department of Biomedical Sciences, Humanitas University, Pieve Emanuele, Milan, Italy}
\IEEEauthorblockA{\IEEEauthorrefmark{4}Radiotherapy and Radiosurgery Department, IRCCS Humanitas Research Hospital, Rozzano, Milan, Italy}
%\{damiano.dei,nicola.lambri\}@cancercenter.humanitas.it, pietro.mancosu@humanitas.it, }
}

\maketitle

\begin{abstract}
A key step in Adaptive Radiation Therapy (ART) workflows is the evaluation of the patient's anatomy at treatment time to ensure the accuracy of the delivery.
To this end, Cone Beam Computerized Tomography (CBCT) is widely used being cost-effective and easy to integrate into the treatment process.
Nonetheless, CBCT images have lower resolution and more artifacts than CT scans, making them less reliable for precise treatment validation.
Moreover, in complex treatments such as Total Marrow and Lymph Node Irradiation (TMLI), where full-body visualization of the patient is critical for accurate dose delivery, the CBCT images are often discontinuous, leaving gaps that could contain relevant anatomical information.
To address these limitations, we propose ARTInp (Adaptive Radiation Therapy Inpainting), a novel deep learning framework combining image inpainting and CBCT-to-CT translation. 
ARTInp employs a dual-network approach: a completion network that fills anatomical gaps in CBCT volumes and a custom Generative Adversarial Network (GAN) to generate high-quality synthetic CT (sCT) images. We trained ARTInp on a dataset of paired CBCT and CT images from the SynthRad 2023 challenge, and the performance achieved on a test set of 18 patients demonstrates its potential for enhancing CBCT-based workflows in radiotherapy.
\end{abstract}


\section{Introduction}
In adaptive radiation therapy (ART), cone beam computed tomography (CBCT) is used to assess any change in the patient's anatomy or tumor volume during the treatment course.
This is particularly important because modern radiotherapy (RT) techniques, such as intensity-modulated radiation therapy (IMRT) and volumetric modulated arc therapy (VMAT), are highly conformal, requiring precise alignment of the patient's anatomy with the treatment plan to deliver the dose accurately.
To this end, CBCT is a cost-effective technology that can be integrated into the Linear Accelerator (LINAC) used to deliver the treatment, allowing the acquisition of real-time images of the patient before each treatment session.
However, when compared to computer tomography (CT), the gold standard imaging modality for radiation therapy (RT) planning, CBCT has a lower image quality and a smaller field of view, which can limit the ability to visualize the patient's anatomy and the regions of interest (ROI) accurately.
This is particularly problematic in complex clinical scenarios, like the conditioning regimes for marrow transplantation~\cite{Passweg2012}, where the ROI spans from head to toe and requires a full view of the patient's anatomy to ensure the adherence of the treatment plan.
A notable example is the total marrow and lymph node irradiation (TMLI)~\cite{Mancosu2019}, a modern treatment option that involves the precise irradiation of the bone marrow and/or the lymph nodes.
As a result, TMLI bears considerably lower late-toxicity issues and is associated with better overall response and clinical outcome~\cite{Cosset1994}. 

Unfortunately, TMLI is still rather underused because of the challenges in its implementation, which include an accurate delineation of the targets and organs at risk (OARs), a demanding planning process to guarantee an accurate dose delivery, and the need for a precise validation of the treatment plan before each session.
In particular, due to the inherent complexity of this treatment, it is extremely important to evaluate the patient's anatomy before each session to ensure the adherence of the treatment plan and to avoid any potential toxicity~\cite{Zhang2015}.
This is typically done through the acquisition of CBCT scans that must be matched with planning CT, i.e., the CT scan used to develop the treatment plan, to make sure that the patient's anatomy has not changed significantly and that the plan is still valid.
However, this process is significantly hindered by the limitations of CBCT, which cannot cover the whole body and has a lower image quality than CT.
\begin{figure}
	\centering
		\includegraphics[scale=0.18]{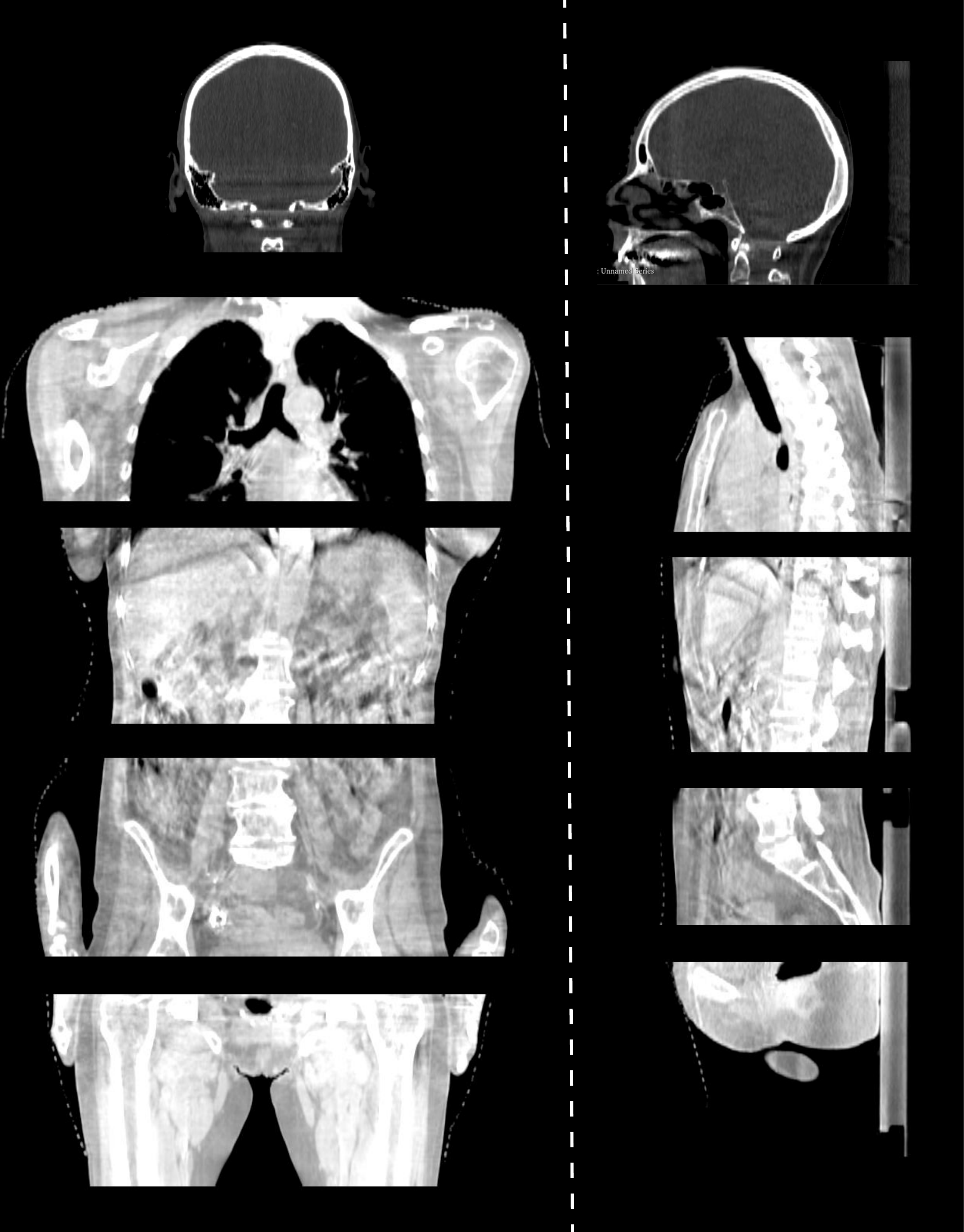}
	\caption{An example of the usage of CBCT before a TMLI session: coronal and sagittal views of several series of CBCT scans combined together to cover the whole body of the patient.}
	\label{fig:gap_overview}
\end{figure}
Fig.~\ref{fig:gap_overview} shows an example of the images that can be obtained with CBCT scans before a TMLI session: it is possible to see the lower image quality and the need to combine several scans to cover the whole body, which results in gaps between the acquisitions.

We aim at addressing the limitations of CBCT in ART workflows for complex treatments like TMLI by proposing a novel deep learning-based framework, called ARTInp (\textbf{A}daptive \textbf{R}adiation \textbf{T}herapy \textbf{Inp}ainting). 
Our framework combines image inpainting to complete the gaps in CBCT images and image translation to enhance the quality of CBCT images, providing a synthetic CT (sCT) image that can be more easily matched with the planning CT to validate the treatment plan.
To the best of our knowledge, ARTInp is one of the first approaches that deal with the problem of generating synthetic images to fill the gaps in CBCT scans acquired during RT procedures.
In this paper, we present a proof-of-concept experiment where we apply ARTInp to a dataset of paired CBCT/CT images of the brain from the SynthRad2023 challenge~\cite{Thummerer2023}, where gaps are artificially introduced in the CBCT images to simulate the clinical scenario described above.
Although this preliminary experiment is set in a much simpler scenario than TMLI or other complex real-world treatments, it demonstrates the potential of ARTInp to enhance the usability of CBCT images and provide a more accurate visualization of the patient's anatomy.
We also hope that, by using a public dataset and a reproducible experimental setup, this work can be a starting point for further research in this direction and the development of more advanced and clinically relevant frameworks.

The paper is organized as follows. In Section~\ref{sec:related}, we review the related works on image inpainting and image translation in the context of medical imaging. In Section~\ref{sec:methods}, we discuss in details the ARTInp framework. In Section~\ref{sec:results}, we describe the experimental setup and present the results. Finally, in Section~\ref{sec:conclusion}, we draw our conclusions and outline the future directions of this research.

\section{Related works}
\label{sec:related}
Synthetic image generation has been a popular research topic since the surge of DL for computer vision tasks in recent years; in the medical field, solutions to generate synthetic CTs are typically based on Generative Adversarial Networks (GANs) or diffusion models \cite{Zhou2023, Wolleb2022}.
Domain transfer and modality generation are two of the most studied tasks, as the necessity for different imaging modalities for the same patient is often considered the optimal approach to correctly define several clinical situations, particularly in RT. 
Among existing works targeting image-to-image translation, MedGAN is one of the first end-to-end approaches \cite{Armanious2018} fully employing DL models successfully; it combines the adversarial loss of GANs with non-adversarial losses to build a framework that allows for PET to CT translation and noisy Magnetic Resonance (MR) refinement in a supervised scenario; from the same authors \cite{armanious2019}, an approach based on cycle consistent GANs (CycleGAN) \cite{zhu2017unpaired} is proposed to tackle a similar problem in an unsupervised setting, which is the most common in the medical field, as it is rather challenging to find paired data for different imaging modalities. 
In \cite{Wolterink2017}, an adapted version of CycleGAN was fed slices from the sagittal plane of different brain MRI modalities, stacked, to generate sCT; in \cite{Jin2019}, a double consistency cycle leads the networks to learn the generation of synthetic MR from CT comparing the synthetic ones with both paired and unpaired images in the discriminator's input; this forces the network to generate a synthetic image which is consistent with both modalities; more recently, another iteration featuring a multi-modal approach to CycleGAN can be found in \cite{Crespi2024}, where multiple branches are added to the generator to develop a multi-modal approach to sCT from MRI acquired with the Dixon method. 
Among works targeting specifically sCT from CBCT, in \cite{Maspero2020}, a CycleGAN approach is used to generate synthetic images that allow for dose calculation from only CBCT through the generation of sCT, focusing on the treatment planning; in \cite{Gao2021} the authors aimed at generating sCT from low-dose CBCT using pix2pix\cite{isola2017image} on paired sets of images and attention-guided GAN\cite{Tang2019} in a CycleGAN setting for the unpaired sets, validating the system with a similar attention to dose calculation; \cite{Deng2023} tried to improve the CycleGAN approach modifying the architecture with an auxiliary chain containing a diversity branch block; \cite{Liu2020} focus on CBCT-based adaptive planning, tackling the sCT generation with CycleGAN modified with self-attention blocks included in the generator; \cite{Gao2023} aims at eliminating streaking artifacts from CBCT to generate once again a sCT on which dose calculation is possible, proposing SARN, a novel architecture, in cascade with CycleGAN and attention-gated CycleGAN; 
in \cite{Fu2024} sCT generation with diffusion models is explored with an energy diffusion model whose de-noising process is performed with a ResUNet with attention blocks and an energy-guided function that retains modality-independent features from the images.

Several efforts can be found in the literature with the objective of inpainting medical images for different purposes. 
In \cite{Pedrosa2025}, an Anatomically-guided Contextual Attention inpainting Network (AnaCAttNet) is guided by anatomical structures to transform pathological tissue of the lungs into healthy tissue.
\cite{Xie2022} proposed a GAN, GatedConv, to inpaint a metal artefact in MRI of the dental implant patients. 
In \cite{Kim2024}, the focus is on inpainting parts of tissues excluded in the field of view of CT acquisition with Globally and Locally Consistent Image Completion (GLIC)\cite{glcic}, obtaining promising results on structural similarity index measure (SSIM). 
An approach based on diffusion models is shown in \cite{durrer2024denoisingdiffusionmodelsinpainting}, with denoising diffusion probabilistic models (DDPM) for the task of transforming tumorous brain tissue in MRI scans into healthy brain tissue. 
To our knowledge, no works have been tackling the inpainting and the sCT generation together.

\section{Methods}
\label{sec:methods}
In this section, details about the dataset and the general implementation of the ARTInp framework are presented, describing the dual architecture, which includes the completion and the translation networks, the dataset and the metrics used for the evaluation. 

\subsection{Dataset}
\label{ssec:dataset}
The dataset used in this work is from the SynthRad 2023 challenge \cite{Thummerer2023}. 
It includes paired CBCT and CT images of the brain, acquired from three Dutch medical centers between 2018 and 2022, with a total of 180 patients (60 patients per center). 
The patients were included regardless of tumor etiology, with age ranging from 3 to 93 years (mean is 65); the images were acquired using different scanner models and manufacturer; these aspects helped in promoting a substantial degree of diversity in the data.
All images were resampled with a voxel spacing of $1 \times 1 \times 1 mm^3$; rigid image registration between the CBCT and the CT was performed using Elastix\footnote{\url{https://elastix.lumc.nl/index.php}}.
Finally, the images were anonymized by defacing the patients and converted in the Nifti format, with each axial slice having a resolution of $256 \times 256$ pixels.
For this work we generated 80\%-10\%-10\% random splits on the patients (144-10-10 patients) for, respectively, training, validation, and test sets; then, we extracted each patient's axial slices and stored them as individual images, encoded in a single channel TIFF format to preserve the entire range of Hounsfield units (HUs).

\subsection{ARTInp Framework}
\label{ssec:ARTInp}
\begin{figure*}[tb]
	\centering
		\includegraphics[scale=0.22]{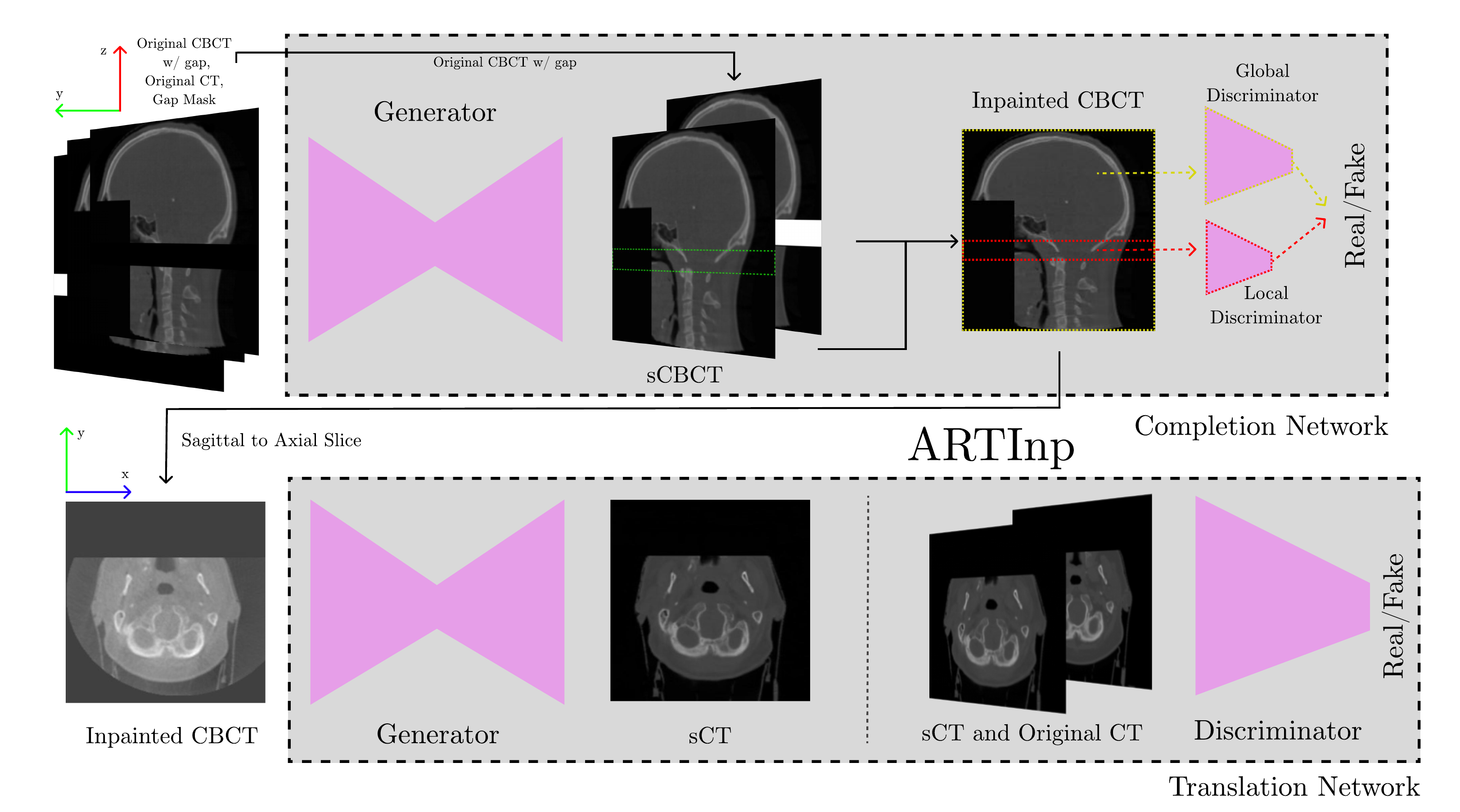}
	\caption{The overview of the ARTInp framework that includes a completion network and a translation network.}
	\label{fig:framwork_inpainting_translation}
\end{figure*}

We propose a novel framework, Adaptive Radiation Therapy Inpainting, dubbed ARTInp, that combines deep learning-based inpainting and CBCT-to-CT translation to enhance CBCT-based workflows in radiotherapy. 
Fig.~\ref{fig:framwork_inpainting_translation} shows the overview of the framework, consisting of two main components: a completion network, that fills the anatomical gaps in the CBCT volumes, and an image translation network, that generates synthetic CT (sCT) images from the inpainted CBCT images.
Both these networks are based on a GAN architecture, which involves a generator for image synthesis a discriminator for evaluating the quality of the generated images.
The completion network has been trained to receive in input a sagittal CBCT slice of the patient with a gap, the corresponding CT slice, and a binary mask indicating the location of the gap in the CBCT slice; the network outputs a \emph{synthetic} CBCT slice with the gap filled (sCBCT in Fig.~\ref{fig:framwork_inpainting_translation}); then, the \emph{inpainted} CBCT is generated by combining the original CBCT with the gap region of 
the generated sCBCT, employing a Poisson image blending~\cite{10.1145/882262.882269} to ensure a smooth transition between the inpainted part and the original ones.
The resulting \emph{inpainted} CBCT slices generated by the completion network are then joined to form the inpainted CBCT volume of the patient, which is used to extract the axial slices that are fed to the translation network.
In fact, the translation network is trained to convert an input \emph{axial} CBCT slice into a corresponding sCT axial slice, allowing to generate a full sCT volume from the \emph{inpainted} CBCT volume. 

%ARTInp employs a dual-network approach: a completion network that fills anatomical gaps in CBCT volumes and a custom Generative Adversarial Network (GAN) to generate high-quality synthetic CT (sCT) images. The inpainting network is trained to fill the gaps in the CBCT images, while the translation network is trained to convert the inpainted CBCT images into sCT images. The inpainting network is trained using sagittal slices of the patient, while the translation network is trained using axial slices. The inpainting network is trained using the CBCT, the corresponding CT, and a binary mask indicating the location of the gap in the CBCT. The translation network is trained using the inpainted CBCT and the corresponding CT. The inpainted CBCT images are converted into axial slices and used as input to the translation network to generate sCT images. The sCT images are then reconstructed into NIfTI files and evaluated using the Mean Squared Error (MSE), Mean Absolute Error (MAE), Peak Signal-to-Noise Ratio (PSNR), and Structural Similarity Index (SSIM) metrics.

\subsection{Completion Network}
\label{ssec:completion}
\begin{figure*}[tb]
	\centering
		\includegraphics[scale=0.36]{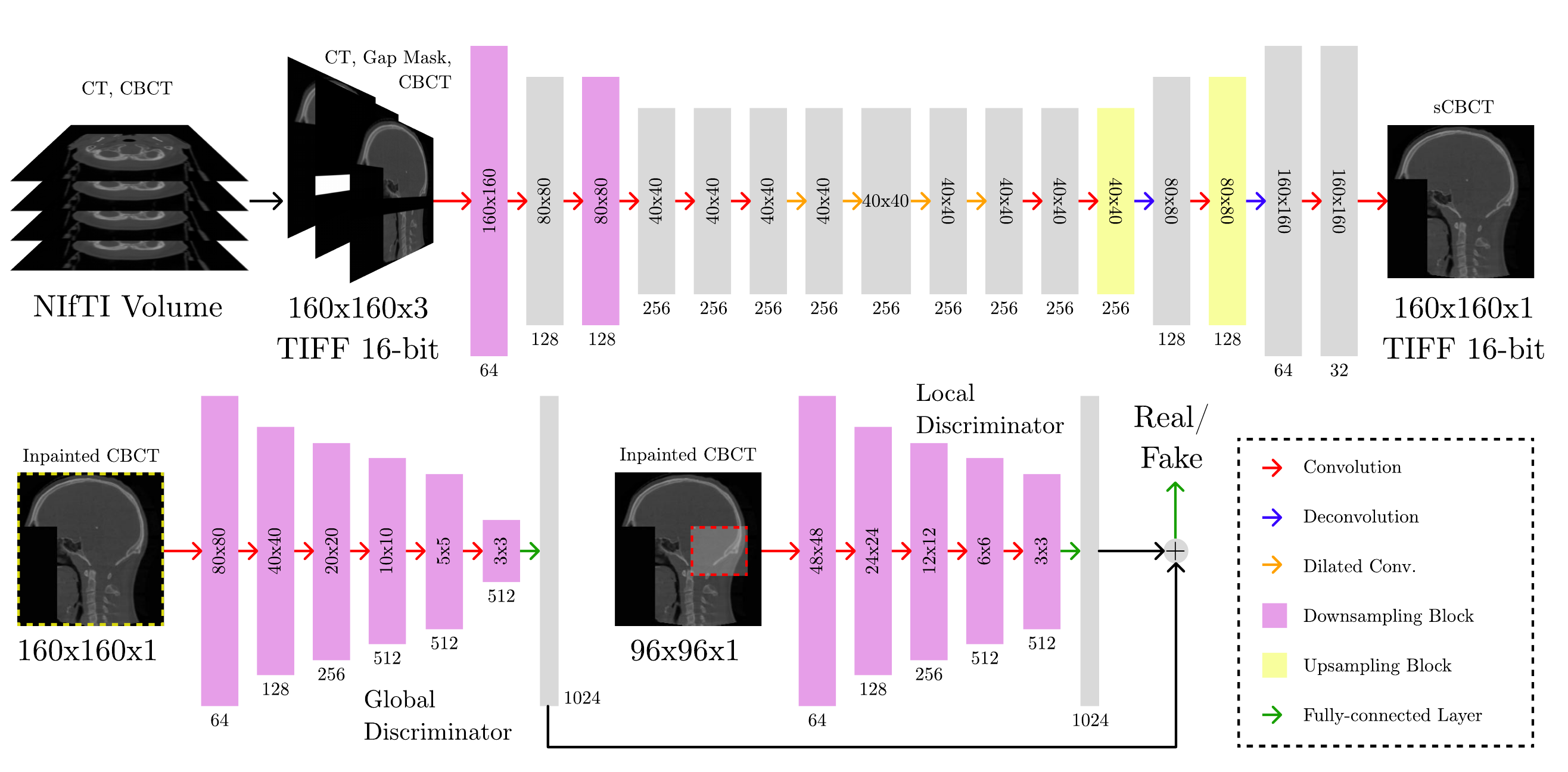}
	\caption{The architecture of the completion network.}
	\label{fig:completion_network}
\end{figure*}
Fig.~\ref{fig:completion_network} shows the architecture of the completion network, which is based on GLCIC~\cite{glcic} and consists of an encoder-decoder structure with convolutional, dilated convolutional, and transposed convolutional layers. 
The encoder-decoder architecture combines convolutional, dilated convolutional, and transposed convolutional layers.
The generator network takes three paired sagittal images of equal size in input: a CT slice, the corresponding CBCT with a gap, and the binary mask of the gap region in the CBCT; the images are 16-bit encoded and have a resolution of $160 \times 160$ pixels.
%There are three images used in the three-channel input of the network, all in the 16-bit range as well and with a 160x160 resolution each one. The sagittal slices of the patient are extracted, both from the CBCT and the CT paired volumes. A gap is generated randomly in the CBCT. and the network takes as input, the sagittal slice of the CT, the sagittal slice of the CBCT with the gap, and, a binary mask with the location of the gap. 
Convolutional layers in the encoding portion of the network have the purpose of extracting features at increasing scale (relatively to the original image) while while dilated convolutions expand the receptive field without reducing resolution; the decoding layers upsample the features, refining details through transposed and standard convolutions in sequence; the final layer produces the grayscale image in output using a $3x3$ convolution and sigmoid activation function. 
To finalise the output, the 16-bit grayscale sCBCT yielded by the network is combined to the input gaped one applying the Poisson image blending to smooth the transition between the inpainted area and the area where the original image was available; the result is a 16-bt grayscale inpainted CBCT; repeating this process for all the slices in the volume and stacking the results, yields the final inpainted CBCT volume. 

The discriminator is a dual-network architecture (a Global Context discriminator and a Local Context discriminator) that evaluates the inpainted CBCT slices at both global and local levels.
The Global Context Discriminator processes the entire inpainted CBCT through $6$ convolutional layers (stride 2, padding 2) with $5\times 5$ filters, each halving spatial dimensions and increasing the number of feature maps; the terminal fully connected layer outputs a 1024-dimensional feature vector capturing the global context.
The Local Context Discriminator focuses on $96\times 96$ patches centered on the inpainted regions; its architecture mimics the Global Discriminator but uses $5$ convolutional layers; it returns a 1024-dimensional vector representing the local context.
The two output vectors are concatenated into a 2048-dimensional vector, processed by a fully connected layer, and passed through a sigmoid activation that yields an estimate of the probability of the image being real.

The training of the completion network consists of three phases: 
\begin{enumerate}
    \item the generator is pre-trained by itself in a supervised fashion;
    \item the discriminators are pre-trained with an semi-adversarial approach where the generator's weights frozen; 
    \item the whole network is trained together in a proper adversarial setting.
\end{enumerate}
The loss function used to train the completion network combines two components: a weighted Mean Squared Error (MSE) loss and an adversarial loss, ensuring both stability and image fidelity: let \( C(x, M_c) \) represent the completion network, where \( x \) is the input image and \( M_c \) is the binary mask indicating the gap region (it has 1's in correspondence of the gap and 0's elsewhere); the MSE loss is defined as follows:
\begin{align}
    \loss_{\text{MSE}}(x, M_c) &= \left\| M_c \odot (C(x, M_c) - x) \right\|_2
\label{eq:mse}
\end{align}
It measures the difference between the CBCT input and the sCBCT output focusing on the regions to be completed.
The adversarial loss combines the discriminators' output in a single term,  \( D(\cdot) \), formulated as a min-max optimization problem: 

\begin{align}
    D(\cdot) = \min_C \max_D \mathbb{E} \Big[ & \log D(x, M_d) + \nonumber \\
    & + \log(1 - D(C(x, M_c), M_c)) \Big]
\label{eq:minmax_gan}
\end{align}

where \( M_d \) is a random mask that determines which parts of the image are considered during the discrimination process, \( D(x, M_d) \) represents the discriminator's output when fed a real CBCT image \( x \) along with the mask \( M_d \), and \( D(C(x, M_c), M_c) \) represents the discriminator's output when fed an sCBCT from the generator.
The discriminator \( D \) aims to maximize its ability to distinguish real and generated images, while the completion network \( C \) attempts to generate images that fool the discriminator.
The final objective function combines the MSE and the adversarial term, balancing them with the parameter $\alpha$, that can be tuned to increase the effect of one term: 

\begin{align}
    \min_C \max_D \mathbb{E} \Big[ &\loss_{\text{MSE}}(x, M_c) + \alpha \big( \log D(x, M_d) \nonumber \\
    & + \log(1 - D(C(x, M_c), M_c)) \big) \Big]
\label{eq:final_loss_completion}
\end{align}

%After the training, the model is used for inference in the test set but this time, the creating of the gaps in the test set is different. In training, the gap was created randomly in any given sagittal slice.
%These holes have the width and height between 48 and 96 pixels and they were positioned randomly within the slice.
%In the testing part, the gap is created patient-wise. It has the width of the sagittal slice and a a height of $2.5cm$, this means that the gap is equivalent to removing a set of entire axial slices from the patient's 3D volume. The gap now represents a missing part of the entire volume ($2.5cm$) and this type of gap is more likely to occur in clinical practice, for example, if you need to combine several CBCT scans over a greater area of the patient and the combination of these scans leaves a gap in between.
%Dividing the total height of the patient's volume based on its slice thickness by the total height of the patient in pixels results in the ratio of millimeters per pixel and thus the $2.5cm$ gap can be calculated individually for each patient.

\subsection{Translation Network}
\begin{figure*}[tb]
	\centering
		\includegraphics[scale=0.36]{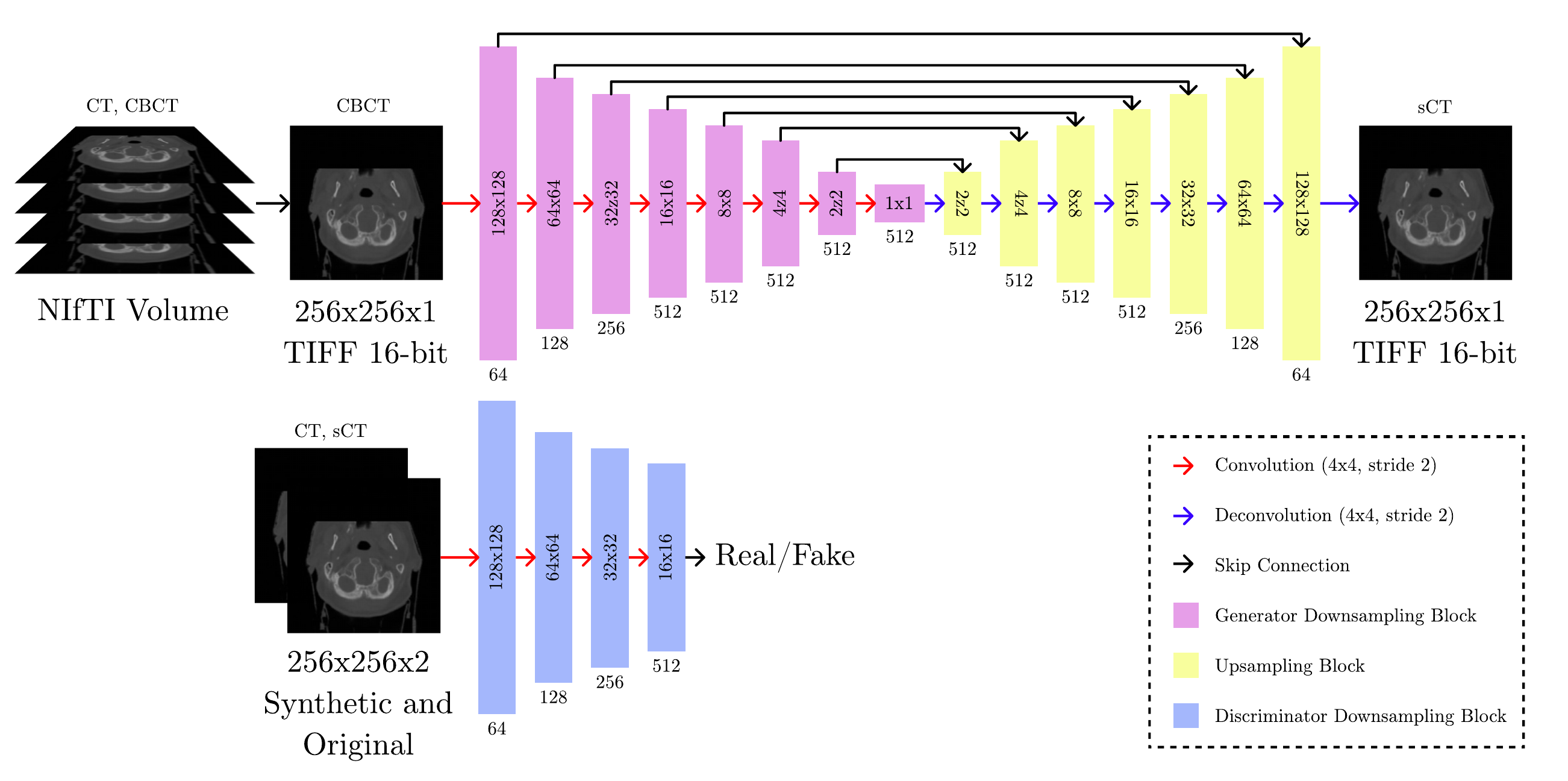}
	\caption{The architecture of the translation network that uses the axial CBCT slices in the 16-bit change and outputs sCT slices.}
	\label{fig:image_translation_network}
\end{figure*}

Fig.~\ref{fig:image_translation_network} shows the architecture of the translation network, based on pix2pix~\cite{isola2017image}; it employs a UNet256 generator and a PatchGAN discriminator.
The generator takes as input a 16-bit $256 \times 256$ axial CBCT slice in input and outputs a 16-bit sCT slice of the same size.
The architecture of the generator consists of an encoder-decoder structure with symmetric layer and skip connections; it features $8$ downsampling blocks comprising a convolutional layer with $4\times4$ filters (stride 2, padding 1), batch normalization, and leaky-ReLU activation; after the bottleneck, the symmetric decoder starts; each decoder block is composed by $4\times4$ transposed convolutions (stride 2, padding 1) to increase the spatial dimensionality, batch normalization, ReLU activation, and skip connections with corresponding encoder layers, that combine high-resolution details from the encoder with the upsampled feature maps, preserving fine-grained details and spatial structure; finally, the outermost block generates the output image through a $4\times4$ transpose convolution (stride 2, padding 1) and a $\tanh$ activation function.
The discriminator is designed to classify $70\times70$ pixel patches of input images as real or fake; by focusing on local patches rather than entire images, it captures fine-grained details efficiently, making it adaptable to various image sizes through its fully convolutional structure.
The architecture of the discriminator consists of a $4\times4$ convolutional layer (stride 2, padding 1) featuring 64 filters, followed by a Leaky ReLU activation; subsequent layers progressively double the number of filters (128, 256, and 512), using the same convolutional structure with batch normalization and leaky-ReLU activation; the final layer generates a single-channel output using a $4\times4$ convolution (stride 1, padding 1), predicting the authenticity for the patches.

The translation network is trained using a loss function composed of two main components: adversarial loss and $\loss_1$ loss.
The adversarial loss ensures that the generator \( G \) produces images that are indistinguishable from real target images \( Y \). To achieve this, a the discriminator \( D \) learns to differentiate between real images \( Y \) and generated images \( G(X) \). The discriminator assigns a probability \( D(y) \) close to \(1\) for real images and \( D(G(x)) \) close to \(0\) for fake images.
The adversarial loss function is formulated as follows:

\begin{align}
    \loss_{\text{GAN}}(G,D) &= \mathbb{E}_{y \sim p_{\text{data}}(y)} 
    \left[ \log D(y) \right]  \nonumber \\ 
    &+ \mathbb{E}_{x \sim p_{\text{data}}(x)} 
    \left[ \log (1 - D(G(x))) \right]
\end{align}

This results in a \emph{min-max} optimization problem:

\begin{equation}
    \min_G \max_D \loss_{\text{GAN}}(G,D)
\end{equation}

While the adversarial loss encourages image fidelity, it does not guarantee that the generated image \( G(x) \) closely matches the corresponding ground truth image \( y \). 
To address this, $\loss_1$ loss enforces pixel-wise similarity:

\begin{align}
    \loss_{1}(G) &= \mathbb{E}_{x,y \sim p_{\text{data}}(x,y)} 
    \left[ \| y - G(x) \|_1 \right]
\end{align}

where \( \| \cdot \|_1 \) denotes the $\loss_1$ norm, which minimizes the absolute differences between the generated image \( G(x) \) and the real image \( y \).

The final objective function is a weighted combination of the adversarial loss and the $\loss_1$ loss:

\begin{align}
    \loss(G,D) &= \loss_{\text{GAN}}(G,D) + \lambda \cdot \loss_1(G)
\end{align}

where \( \lambda \) is a hyperparameter that controls the trade-off between adversarial realism and structural similarity. The full optimization objective remains:

\begin{align}
    \arg \min_G \max_D & \loss_{\text{GAN}}(G,D) + \lambda \cdot \loss_1(G)
\end{align}

This formulation ensures that the generated images are both visually realistic and structurally aligned with the corresponding ground truth images.
%The image translating training ran for 60 epochs, using a batch size of 1 in the Kaggle platform.

\subsection{Evaluation Metrics}
To evaluate the quality of sCT images generated by our models, we used three metrics: 
\begin{itemize}
    \item Mean Percentage Absolute Error (MAE\%): measures the average absolute difference between the pixel values of the generated images and the ground truth images as a percentage of the overall pixel value range;
    \begin{equation*}
        \text{MAE\%} = \frac{100}{N \cdot R} \sum_{i=1}^{N} \lvert I_i - \hat{I}_i \rvert \\[10pt]
    \end{equation*}
    
    \item Peak Signal-to-Noise Ratio (PSNR): quantifies the ratio between the maximum possible power of the image and the power of the noise that affects the image;
    \begin{equation*}
        \text{PSNR} = 10 \cdot \log_{10} \left( \frac{I_\text{MAX}^2}{\frac{1}{N} \sum_{i=1}^{N} (I_i - \hat{I}_i)^2} \right) \\[10pt]
    \end{equation*}
    \item Structural Similarity Index (SSIM).
    \begin{equation*}
        \text{SSIM}(I, \hat{I}) = \frac{(2 \mu_I \mu_{\hat{I}} + C_1)(2 \sigma_{I\hat{I}} + C_2)}{(\mu_I^2 + \mu_{\hat{I}}^2 + C_1)(\sigma_I^2 + \sigma_{\hat{I}}^2 + C_2)}
    \end{equation*}
\end{itemize}

In the formulas above, $N$ is the number of pixels in the image, $I_i$ $\hat{I}_i$ are the pixel values of the original and generated image respectively, $R$ is the range of pixel values in the dataset, $I_\text{MAX}$ represents the maximum possible pixel value, $\mu_I$ and $\mu_{\hat{I}}$ are the mean values of the original and generated images, $\sigma_I^2$ and $\sigma_{\hat{I}}^2$ are their variances,$\sigma_{I\hat{I}}$ is the covariance between the two images, $C_1$ and $C_2$ are small constants to stabilize the division.

All these metrics are computed slice by slice considering only the region of the image that contains the patient, such that the background of the image does not affect the calculation.

\section{Results}
\label{sec:results}
\subsection{Experimental Design}
To evaluate the performance of the proposed ARTInp framework, we conducted two experiments: an evaluation of the performance of the translation network on the task of converting CBCT to sCT in the absence of the completion network; a measure the performance of the whole ARTInp framework.

In the first experiment, the translation network was trained using the CBCT and CT images pairs from the 144 patients of the training set (see Section \ref{ssec:dataset}) for 60 epochs (around 1800000 iterations) using an Adam~\cite{Kingma2015} optimizer with a learning rate of $2\cdot10^{-4}$ and a batch size of 1. 
%The validation set of 10 patients was used to tune the training parameters, that was selected as follows. 
A checkpoint was saved every 5 epochs and the model with the best performance on the validation set was selected. 

In the second experiment, we trained the completion network by generating artificial gaps in the same images used for the previous experiment. 
In particular, we created random gaps in the CBCT slices included in the training set by removing vertical strips of the images with a random location and a pixel width $w$ randomly sampled from a uniform distribution $\mathcal{w}\sim\mathcal{U}(48, 96)$.
The resulting CBCT images with artificial gaps were used, along with a binary mask of the generated gaps and the corresponding original CT images, to train the completion network.
These artificial gaps mimic the ones that would be found in a real clinical scenario, having similar width and positioning. 
The three training phases described in Section~\ref{ssec:completion} where implemented as follows: 
\begin{enumerate}
    \item in the first phase the generator network is trained using only the L1 loss for 180000 iterations;
    \item in the second phase, the generator network's weights are frozen and only the discriminator network is trained for 20000 iterations;
    \item in the third phase, both the generator and the discriminator networks are trained together for 620000 iterations.
\end{enumerate}

The batch size was 1 and the ADADELTA optimization algorithm~\cite{zeiler2012adadelta} was used; a checkpoint was saved every 2000 iterations and the model with the best performance on the validation set was selected.

The evaluation of the models trained in the two experiments described above was performed on the test set of 18 patients.
To evaluate the performance of the translation network, we tested it on the 3456 CBCT axial slices included in the test set and computed the MAE\%, PSNR, and SSIM metrics.
Then, we combined the translation network with completion network trained in the second experiments to evaluate the performance of the whole ARTInp framework.
We created artificial gaps in the CBCT slices included in the test set, similarly to what we did in the training set; however, in this case, the location and the width of the gaps were sampled patient-wise, i.e., we removed the same portion in all the slices of the CBCT volume of a patient, to adhere as much as possible to the actual clinical situation.
The inpainted CBCT images of the test set generated with the completion network were then used to generate the sCT images through the translation network.
Finally, we computed the MAE\%, PSNR, and SSIM metrics to evaluate the quality of the images generated by the ARTInp framework with respect to the original CT images.
%After the training, the model is used for inference in the test set but this time, the creating of the gaps in the test set is different. In training, the gap was created randomly in any given sagittal slice.
%These holes have the width and height between 48 and 96 pixels and they were positioned randomly within the slice.
%In the testing part, the gap is created patient-wise. It has the width of the sagittal slice and a a height of $2.5cm$, this means that the gap is equivalent to removing a set of entire axial slices from the patient's 3D volume. The gap now represents a missing part of the entire volume ($2.5cm$) and this type of gap is more likely to occur in clinical practice, for example, if you need to combine several CBCT scans over a greater area of the patient and the combination of these scans leaves a gap in between.
%Dividing the total height of the patient's volume based on its slice thickness by the total height of the patient in pixels results in the ratio of millimeters per pixel and thus the $2.5cm$ gap can be calculated individually for each patient.

\subsection{Results}
\begin{table}
	\centering
	\begin{tabular}{lccc}
	MODEL                  & MAE\%         & PSNR           & SSIM            \\ \hline
	
	ARTInp w/o Completion         & 2.22   ± 1.35 & 27.77   ± 2.74 & 0.791   ± 0.082 \\
	ARTInp & 2.44   ± 1.25 & 26.84   ± 2.53 & 0.768   ± 0.079
	\end{tabular}
	\label{tab:set1_results}
	\end{table}	
Table~\ref{tab:set1_results} shows the performance achieved by ARTInp models on the test set.
In the table, \emph{ARTInp w/o Completion} refers to the model that only includes the translation network and does not perform any inpainting, while \emph{ARTInp} refers to the model that includes both the translation and the completion networks.
Both models achieve an MAE\% below 2.5\%, corresponding to average values under 100 HU, which is consistent with what has been observed in the literature for the brain area~\cite{edmund_review_2017}.
The PSNR around 27dB and an SSIM below 0.8 suggest the presence of artifacts and distortions compared to the real images. 
However, the introduction of gaps and the use of the completion network do not seem to drastically degrade the performance, keeping it at promising values for the practical application of ARTInp.

\begin{figure}
	\centering
		\includegraphics[scale=0.11]{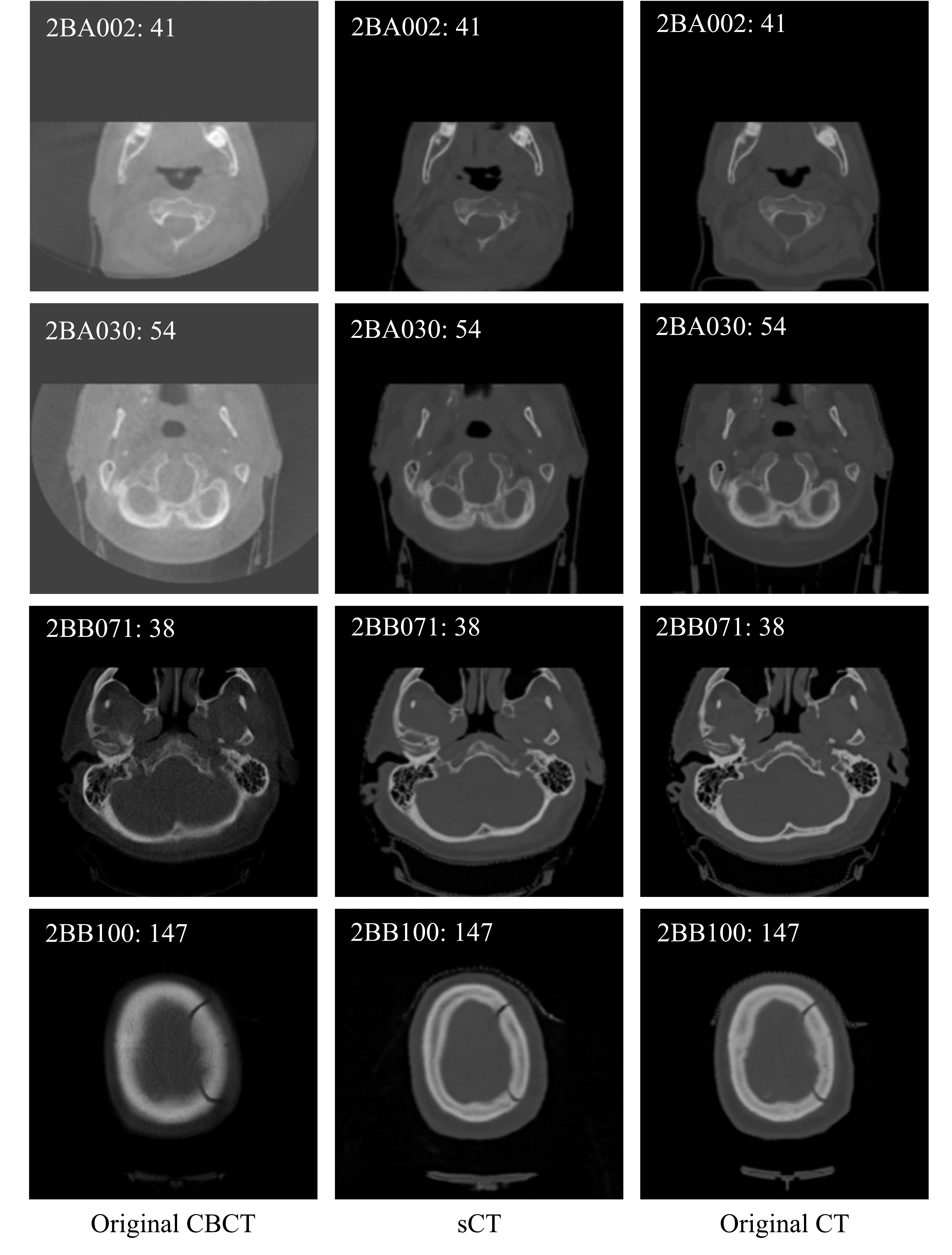}
	\caption{Examples of the sCT generation.}
	\label{fig:8bit_pix2pix}
\end{figure}

\begin{figure}
	\centering
		\includegraphics[scale=0.11]{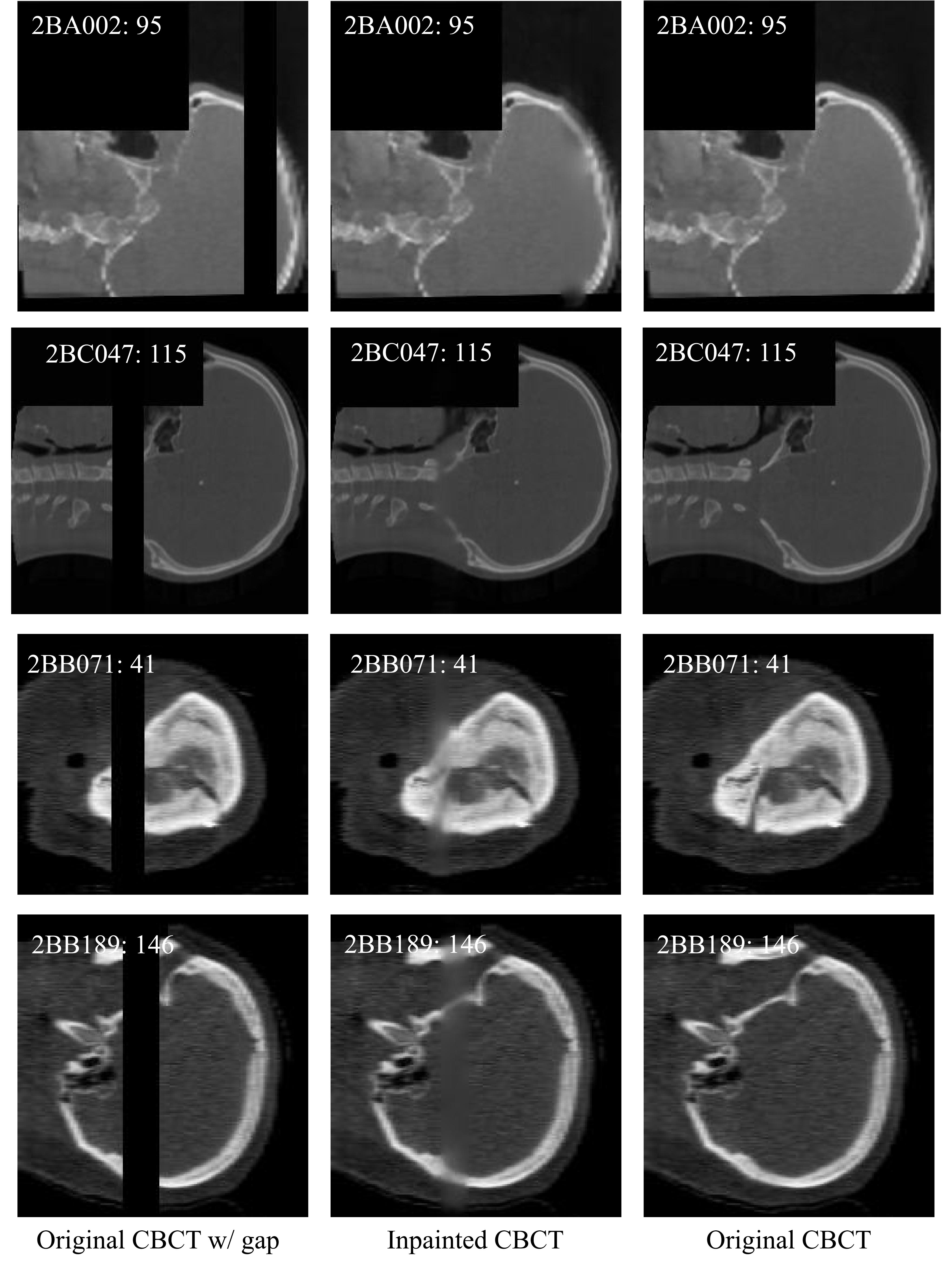}
	\caption{Examples of the inpainted CBCT. The images contain the original CBCT with a gap, the inpainted CBCT, and, the original CBCT.}
	\label{fig_gap_ct}
\end{figure}

We also present some qualitative results of the sCT generation and the inpainted CBCT in Fig.~\ref{fig:8bit_pix2pix} and Fig.~\ref{fig_gap_ct}, respectively.
The sCT images generated by the translation network are shown in the second column of Fig.~\ref{fig:8bit_pix2pix}, while the original CBCT image and the original CT images are respectively in the first and third columns for comparison.
The inpainted CBCT images are shown in the second column of Fig.~\ref{fig_gap_ct}, while the original CBCT images with gaps and the original CBCT images are respectively in the first and third columns for comparison.
The examples show how the sCT images generated are generally able to capture quite well the features of the original CT images, even if some artifacts are present.
In Figure \ref{fig_gap_ct}, the inpainted CBCT are shown in the second column, while the original CBCT is presented on the third column for comparison. 
It is possible to observe that the inpainting process is able to reconstruct the missing parts of the CBCT images, even if some artifacts are present, such as blurriness and noise.
In particular, in the first and fourth example, the bone of the skull seems blurry in the gap location and the tissue inside the skull has significantly more noise than the original CBCT, although the structure's shape still is close to the original. 
In the second example, the atlas vertebrae is inpainted successfully along with the beginning of the skull bone. 
The third example is a region on the side of the patient's skull and it can be seen that the suture that separates the parietal and the temporal bone is slightly blurrier in the inpainted CBCT.

\section{Conclusion}
\label{sec:conclusion}
We have presented ARTInp, a novel framework for inpainting and translating CBCT images into synthetic CT images, consisting of two networks: a completion network that fills the gaps in CBCT images and a translation network that generates synthetic CT images from the inpainted CBCT images. 
We designed ARTInp to enhance the application of CBCT images in the ART workflow, particularly in complex treatments such as TMLI, where the target area spans on the whole body of the patient and an analysis over a large volume is crucial.

To evaluate the potential of ARTInp, we trained it on a dataset of paired CBCT and CT images made available for the SynthRad 2023 challenge \cite{Thummerer2023}.
In particular, the completion network was trained to inpaint CBCT images with artificial gaps, while the translation network was trained to generate synthetic CT images from paired CBCT images.
At inference time, the completion network was used to inpaint the CBCT images with artificial gaps, which were then fed to the translation network to generate the synthetic CT images, which were compared to the original CT images.
The performance of ARTInp is promising, with an MAE\% below 2.5\% and a PSNR around 27dB, suggesting that the quality of the synthetic CT images generated is similar to the one achieved by models with similar target tasks in the literature~\cite{edmund_review_2017}, but no other works presented the full pipeline, including the inpainting. 
In particular, when CBCT images are inpainted before generating the synthetic CT images, the final performance is only slightly degraded, showing that the inpainting process is effective.
Nevertheless, the SSIM values are below 0.8, and a qualitative analysis of the generated images shows the presence of artifacts and distortions such as blurriness and noise, limiting so far the possibility to apply the framework in clinical processes. 
Overall, the results suggest that ARTInp has the potential to enhance CBCT-based workflows in radiotherapy and is worth further investigation.

Future work will focus on investigating additional training strategies, such as training both networks together, as well as using more complex architectures, such as 3D networks and diffusion models~\cite{Wolleb2022}, to improve the quality of the generated images. 
Moreover, we plan to evaluate the performance of ARTInp on a specific clinical setting, such as TMLI, to assess its potential for enhancing the ART workflow in radiotherapy.

%\section*{Acknowledgment}
%This work is part of the AuToMI project funded by the Italian Ministry of Health (Rome, Italy; grant number: GR-2019-12370739).

\bibliographystyle{ieeetr} 
%\linespread{0.9}
\balance
\bibliography{myrefs.bib}

\end{document}